# Magnetic Frustration and Spin-Charge Separation in 2D Strongly Correlated Electron Systems

W. O. Putikka

*National High Magnetic Field Laboratory, Florida State University,
Tallahassee, FL 32306-4005*

**Abstract**

I propose that the ground state of the 2D $t$-$J$ model near half-filling with $J/t \sim 1/3$ has both ferromagnetic and antiferromagnetic fluctuations leading to a magnetically frustrated ground state. I further argue that the frustrated state is spin-charge separated to account for the observed behavior of the equal time spin and charge correlation functions.

Attempts to understand high $T_c$ superconductors have centered around the 2D copper oxide planes found in these materials. One would like to know which features of the planes in the normal state are essential to achieving high $T_c$'s. Considerable effort has gone into understanding strongly correlated electrons on a square lattice[1]. In particular, the single band 2D Hubbard and $t$-$J$ models near half-filling have been studied intensively. The data discussed in this paper are from high temperature expansions[2–4] for the 2D $t$-$J$ model, concentrating on the optimal doping range $n \approx 0.80$–$0.85$ and $J/t \sim 1/3$. I consider a scenario for the planes in which strong correlations, two-dimensionality and the bipartite nature of the square lattice all play important roles.

For models with repulsive interactions ($U > 0$) the most likely instabilities are magnetic. It is well established[5] that the 2D Hubbard model at half-filling and $T = 0$ is an ordered antiferromagnet (AF) with ordering wavevector $\vec{Q} = (\pi, \pi)$. Away from half-filling AF order disappears rapidly for the Hubbard or $t$-$J$ model and is no longer present[2] for $n \lesssim 0.95$, $J/t \sim 1/3$. For the uniform susceptibility $\chi_0(T, n)$ with similar model parameters and $T < J$ different behavior is observed. Initially upon doping $\chi_0(T<J, n)$ increases, going through a maximum[2] at $n \approx 0.80 - 0.85$. This maximum is larger for smaller $J/t$, but remains at the same density. The size of the maximum is determined by the proximity to a ferromagnetic (FM) region[3] at smaller $J/t$ which has its greatest extent in $J/t$ for $n \approx 0.80 - 0.85$. The FM region for $J/t > 0$ is not fully polarized and most likely is part of a second order transition from the fully polarized state at $J/t < 0$ to a paramagnet at larger $J/t$. Having a preferred density for FM behavior is due to two effects which act to suppress FM fluctuations. Near half-filling for $J/t > 0$ AF fluctuations are favored, while at low densities the strong correlation effects which produce magnetic behavior are reduced (Kanamori paramagnetism[6]).

The equal time correlation functions[2] for the 2D $t$-$J$ model provide further information on the spin and charge degrees of freedom for $n \approx 0.80 - 0.85$. The spin correlation function $S(\vec{q})$ (charge correlation function $N(\vec{q})$) has $2\vec{k}_F$ ($2\vec{k}_F^{SF}$) as a characteristic wavevector, where $\vec{k}_F$ ($\vec{k}_F^{SF}$) is the Fermi wavevector for the non-interacting tight binding (spinless fermion) model on a square lattice. In 2D $\vec{k}_F$ and $\vec{k}_F^{SF}$ are *incommensurate* and related by a doubling of the number of occupied $\vec{k}$-states in the Brillouin zone.

The AF and FM correlation lengths $\xi$ can be calculated from the spin susceptibility $\chi(\vec{q}, 0)$ by

$$\chi(\vec{q}, 0) = \chi(\vec{Q}) \left[ 1 - K_{AF} \left( (q_x - \pi)^2 + (q_y - \pi)^2 \right) + \cdots \right], \qquad (1)$$

where $\vec{Q} = (\pi, \pi)$ and if $K_{AF} > 0$ we have $K_{AF} = \xi_{AF}^2$. For the FM case $K_{FM}$ and $\xi_{FM}$ are defined by a similar expansion around $\vec{q} = (0, 0)$. Near half-filling we expect $\xi_{AF}/a \gg 1$ where $a$ is the lattice spacing, with $K_{FM} < 0$. This behavior is shown as curves a and b in Fig. 1. However, upon further doping to $n = 0.80$ a different picture emerges. Now $\xi_{AF}$ is *small*, $\xi_{AF}/a \sim 0.3$ and $K_{FM}$, while still negative, is becoming closer to a proper correlation length at low $T$, shown as curves c and d in Fig. 1. Keeping $n = 0.80$ there is a crossover in the low

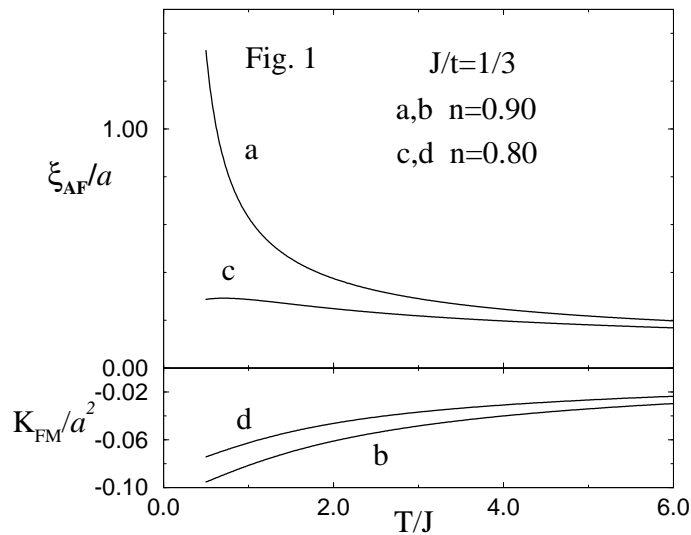

$T$ behavior of $K_{AF}$ and $K_{FM}$ for $J/t \lesssim 0.2$. Now at low $T$, $\xi_{FM}$ exists as a well defined correlation length, while $\xi_{AF}$ does not exist as shown in Fig. 2. This suggests that there is a region around $n \approx 0.80$, $J/t \sim 1/3$ where the magnitude

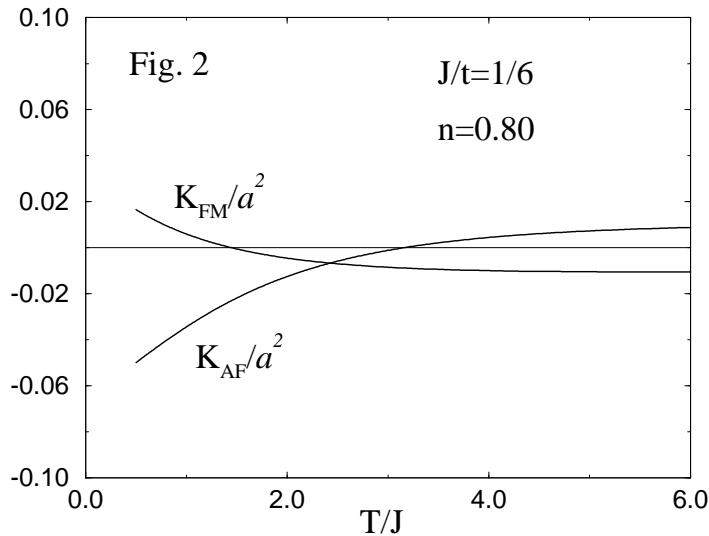

of $\chi(\vec{q})$ is enhanced, but $\chi(\vec{q})$ is also rather flat as a function of $\vec{q}$.

The above observations for the 2D $t$-$J$ model suggest that for $n \approx 0.80$, $J/t \sim 1/3$ both AF and FM fluctuations are important, leading to a magnetically frustrated[7] ground state. Further, the characteristic wavevectors observed for $S(\vec{q})$ and $N(\vec{q})$ in the same parameter region suggest that the frustrated magnetic state leads to distinct 2D momentum distributions for the spin and charge degrees of freedom. This is a form of spin-charge separation unique to 2D. Another way to view the frustrated state is as resulting from competition between tendencies towards long range AF and FM order with an ordered state at $T = 0$ prevented by an exact degeneracy between the two types of fluctuations. The exact degeneracy is plausible in 2D for Heisenberg spins, which by the Mermin-Wagner theorem[8] have $T_c = T_N = 0$. This is not true in higher dimensions where in general $T_c \neq T_N$ and the ground state is likely to be the ordered state with the highest transition temperature. Only by extreme fine tuning of parameters in the model is equality likely in higher dimensions.

With the point of view expressed here the analogy of 2D to 1D strongly correlated models is indirect, resting on the idea of (different) degenerate, competing instabilities. In 1D there are competing nesting instabilities which must be treated on an equal footing to find the correct ground state, resulting in the Tomonaga–Luttinger liquid with separate spin and charge excitations[9]. In 2D the competing instabilities are AF and FM, but now if $\chi(\vec{q})$ diverges as $T \to 0$ for a fixed $\vec{q}$ there is long range order. Thus one distinguishing feature for a spin-charge separated state in 2D should be an enhanced, relatively flat $\chi(\vec{q})$. The nature of the competing instabilities in 2D also restricts the possible spin-charge separated state to densities near half-filling, unlike 1D where the nesting instabilities exist for all densities. The phenomenology of spin-charge separation[10] is likely common to 1D and 2D strongly correlated systems (at least for a limited range of parameters in 2D) though the underlying mechanisms are distinct.

The presence of frustration near half-filling could be an intrinsic source of the sign problem in Monte Carlo calculations for the 2D Hubbard model. While the sign problem can be affected by different Hubbard–Stratonovich decompositions, preventing a rigorous conclusion, the correspondence of the parameter ranges for the sign problem and magnetic frustration is striking. As shown by Scalapino[11] the sign problem is most extreme for $n \sim 0.80 - 0.85$ and gets worse as $U/t$ is increased. Both behaviors match up very well with the parameter range for the strongest magnetic frustration.

In conclusion, I have discussed the properties of the 2D $t$-$J$ model (or large $U$ Hubbard model) for $n \approx 0.80 - 0.85$ and $J/t \sim 1/3$. I propose that the 2D $t$-$J$ model for this parameter range is magnetically frustrated between exactly degenerate, competing AF and FM instabilities, leading to a spin-charge separated ground state distinct to 2D. This state requires the distinguishing features of the copper oxide planes: a single band, strong, repulsive correlations, two dimensionality, doping near half-filling and a bipartite lattice. The behavior of FM in the 2D $t$-$J$ model also offers an explanation for the optimal doping range $n \approx 0.80 - 0.85$ observed in all high $T_c$ superconductors.

This work was supported by the National High Magnetic Field Laboratory at Florida State University and by NSF Grant No. DMR-9222682.

# References

[1] K. S. Bedell, *et al.*, eds., *High Temperature Superconductivity* (Addison-Wesley, Redwood City, CA, 1990); K. S. Bedell, *et al.*, eds., *Strongly Correlated Electronic Materials* (Addison-Wesley, Reading, MA, 1994).

[2] R. R. P. Singh and R. L. Glenister, *Phys. Rev.* **B46** (1992) 11871; W. O. Putikka, *et al.*, *Phys. Rev. Lett.* **73** (1994) 170.

[3] W. O. Putikka, *et al.*, *Phys. Rev. Lett.* **69** (1992) 2288.

[4] A. Sokol, *et al.*, *Phys. Rev. Lett.* **72** (1994) 1549.

[5] E. Manousakis, *Rev. Mod. Phys.* **63** (1991) 1.

[6] J. Kanamori, *Prog. Theor. Phys.* **30** (1963) 276.

[7] T. M. Rice in Ref. 1, *Strongly Correlated Electronic Materials*, 494.

[8] N. D. Mermin and H. Wagner, *Phys. Rev. Lett.* **22** (1966) 1133.

[9] J. Solyom, *Adv. Phys.* **28** (1979) 201.

[10] P. W. Anderson, *Phys. Rev. Lett.* **64** (1990) 1839.

[11] D. J. Scalapino in Ref. 1, *High Temperature Superconductivity*, 314.